\documentclass[11pt, a4paper]{article}
\usepackage{jinstpub}

\usepackage{here}
\usepackage[T1]{fontenc}
\usepackage{tgcursor}
\usepackage{verbatim}
\usepackage{lineno}
\title{Impact of the hole orientation of asymmetric GEM foils on the performance of single and triple GEM detectors}

\author[a]{O.~Bouhali,}
\author[b]{K.~Hoepfner,} 
\author[b]{F.~Ivone,} 
\author[d]{T.~Kamon,}
\author[b]{H.~Keller,}
\author[a,1]{S.~Malhotra %
\note{Corresponding author.}}
\author[c]{and B.Al~Rashdi}

\affiliation[a]{Science Program, Texas A$\&$M University at Qatar,\\Education City, Doha, Qatar} %
\affiliation[b]{Physics Department, 3. Phys. Inst. A, RWTH Aachen University,\\Otto-Blumenthal-Str., Aachen, Germany} %
\affiliation[c]{Department of Physics, College of Science, Sultan Qaboos University,\\Al Khoudh, Muscat, Oman}
\affiliation[d]{Physics \& Astronomy, Mitchell Physics Building, Texas A$\&$M University,\\578 University Drive, College Station, Texas} %

\emailAdd{shivali.malhotra@cern.ch}

\abstract{The Gas Electron Multiplier (GEM) foil is an amplification stage that has been introduced to overcome the problem of discharges observed in gaseous detectors. There are two major production techniques of GEM foils: double-mask and single-mask etching. Despite being an effective method, an asymmetry is observed between the top and bottom diameters of GEM holes in single mask technique compared to double mask one. In this paper we describe extensive simulations and measurements to study this hole asymmetry and its effect on the performance of GEM based detectors. The experimental data is collected using GEM foils of various hole geometries and orientations. In simulations, the same dimensions are used to study the properties of the detector. Simulations are performed with the Garfield$^{++}$ simulation package along with ANSYS for creating the geometry of the GEM foils as well as the triple-GEM detector and the meshing needed for the field calculations. The simulation results match the observations from experimental studies. The gains measured with single and triple-GEM detectors are lower if asymmetric foils are oriented with the smaller diameters towards the readout plane. Detailed simulation of the amplification and collection steps indicates that the lower gain is attributed to a loss of electrons at the GEM3 foil for the first time.
}


\begin{document}
\maketitle
\flushbottom

\section{Introduction}
\label{sec:int}

The last two decades have witnessed the emergence of several Micro-Pattern Gas Detectors (MPGD)~\cite{Sauli,Micromegas} that have been used in various science areas such as: high energy physics, astronomy and medical physics~\cite{Sauli2}. MPGDs provide a number of features that make them attractive: high gain, excellent detection efficiency, spatial resolution, high rate capability and robustness in addition to their relatively low cost. These detectors have been extensively studied in the past both through measurements and simulations. 

An important recent application of MPGD is within the future upgrade of the Large Hadron Collider (LHC). The operation plan of LHC calls for a series of long periods of data taking (Run 1, Run 2, Run3, Run 4 etc.) interleaved with periods of the accelerator complex shutdowns (so called “Long Shutdown” LS1, LS2, LS3 and LS4). Consequently the associated experiments, such as Compact Muon Solenoid (CMS)~\cite{CMS}, will undergo a series of upgrades to preserve detection efficiency, timing and spatial resolutions and optimal background rejection at the highest luminosity~\cite{CMStdr}. 

Gas Electron Multiplier (GEM) technology~\cite{Gemtdr} is adopted by CMS for installing additional muon chambers in the endcap region~\cite{Gemtdr, MuonTDR}, where harsh environmental background hit rates are expected. 
A challenge in GEM detectors for CMS and future experiments is to design, optimize and assemble these large GEM foils with double mask (DM) or single mask (SM) techniques~\cite{Abbaneo2,CMSgem2} in particular if combined in the same detector system. These foil techniques introduce hole-size asymmetry between the top and bottom of a GEM hole diameter, leading to variation in gas gain. A proper understanding of the underlying correlation was not known for a long time. 

In this paper, we report an extensive study of gas gain variation due to the hole asymmetry in the GEM foils used for GEM detectors. In section~\ref{sec:setup}, we describe the simulation set-up and in section~\ref{sec:geometry} the detector geometry by focusing on the hole dimensions we are considering for single and triple-GEM detectors. Section~\ref{sec:ExpMeas} provides the details corresponding to the experimental set-up used for performing measurements. Finally the results from both the simulations and measurements are presented and discussed in section~\ref{sec:results}.

%%%%%%%%%%%%%%%%%%%%%%%%%%%%%%%%%%%%%%%%%%%%%%%%%%%%%%%%%%%%

\section{Simulation set-up}
\label{sec:setup}

\begin{figure}[hbtp]
  \begin{center}
    \resizebox{9cm}{!}{\includegraphics{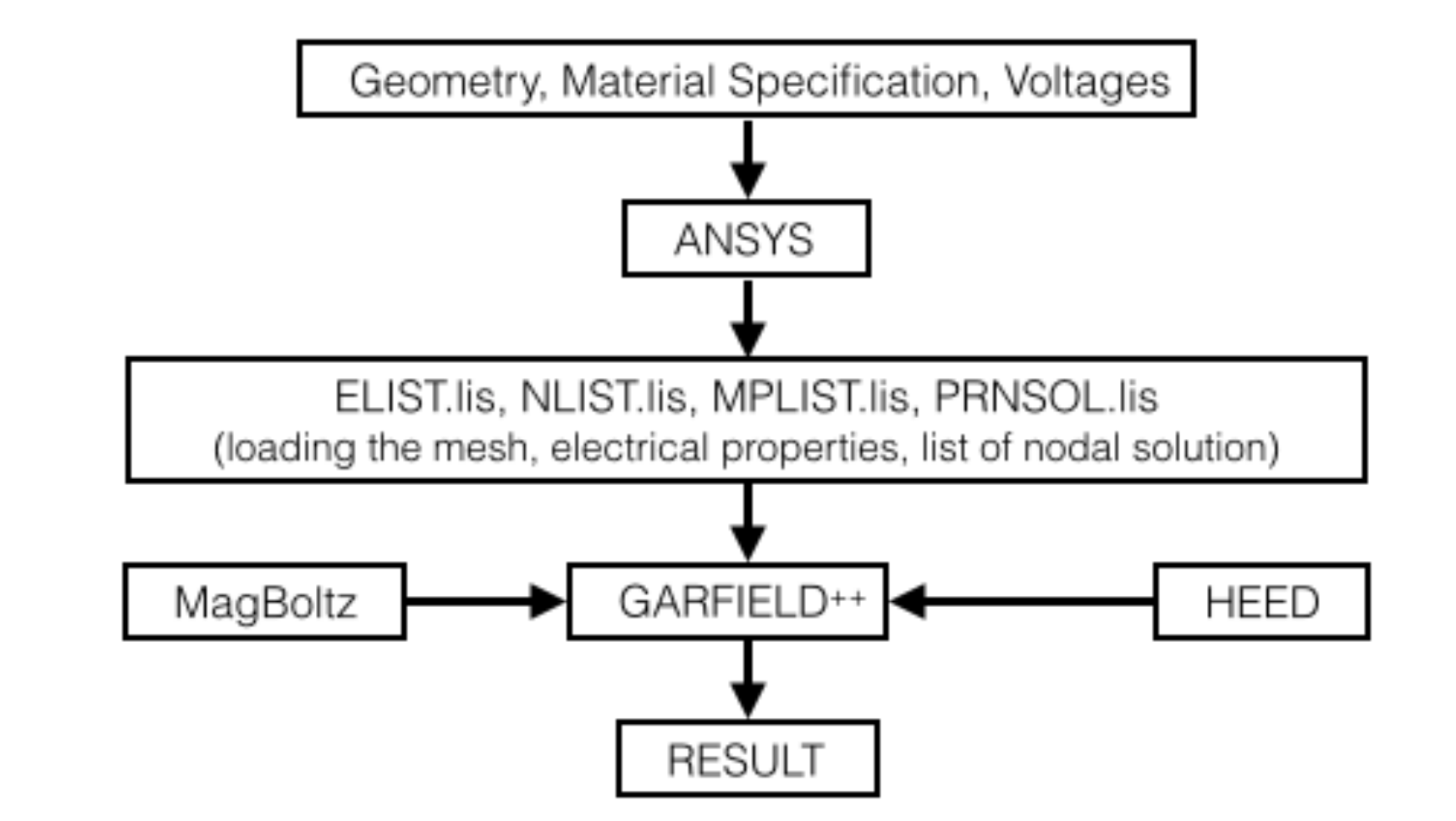}}
    \caption{Simulation workflow for GEM detector performance study.}
    \label{fig:simflow}
  \end{center}
\end{figure}

The simulation is based on Garfield$^{++}$, a C$^{++}$ version of Garfield~\cite{Veenhof}. The latter has been developed during the eighties to simulate drift chambers. Since then it has evolved to cover the simulation of ionization gas and semiconductor based detectors.
For complex geometries the electric field and detector geometry can be computed with an external software such as ANSYS~\cite{ANSYS}, COMSOL~\cite{COMSOL}, neBEM~\cite{nebem}, etc. The field map generated from any of these software is used as input for Garfield$^{++}$. It is further interfaced with two other codes: MagBoltz~\cite{Magboltz} which numerically solves the Boltzmann transport equation for electrons in gas under the influence of electric and magnetic fields, and HEED~\cite{HEED} that simulates the ionization of gas molecules by an incident charged particle. A descriptive diagram for the simulation workflow is shown in figure~\ref{fig:simflow}. 

In this work, we first generate the electric field map using ANSYS which is a computational fluid dynamics software based on numerical finite element analysis. A dedicated ANSYS script is written defining the hole geometry for each GEM foil, geometry used for triple-GEM detector, material specifications and voltages provided at the different electrodes. ANSYS then generates a set of files containing a list of detector elements with a table detailing their electrical properties, a list of nodes and their positions in space and the calculated potentials at each of the nodes. The combination of ANSYS with Garfield$^{++}$ is a very useful tool that has been used previously to simulate a large number of gas based detectors~\cite{Bachmann1,Nim2016}. As the simulation is very cumbersome and time consuming, in particular when the detector geometry is complex or large (e.g. triple-GEM detector) or when a high gas gain is required, we have used the parallel and optimized versions of Garfield$^{++}$ introduced in~\cite{PGARFIELD}.

%%%%%%%%%%%%%%%%%%%%%%%%%%%%%%%%%%%%%%%%%%%%%%%%%%%%%%%%%%%%

\section{GEM detector and foil geometry}
\label{sec:geometry}

In this section we provide a detail description of the geometries considered for the measurements and the simulations. First we will describe the different configurations of the GEM foils which were used and then introduce the different GEM detector geometries which were studied in detail. The configurations of the GEM foils are distinguished based on the geometry of the holes. It also explains how these various configurations are used to build triple-GEM detectors along with their operating voltages.

\subsection{GEM detector geometry}
\label{subsec:GEMgeometry}

\begin{figure}[hbtp]
  \resizebox{7.25cm}{5.9cm}{\includegraphics{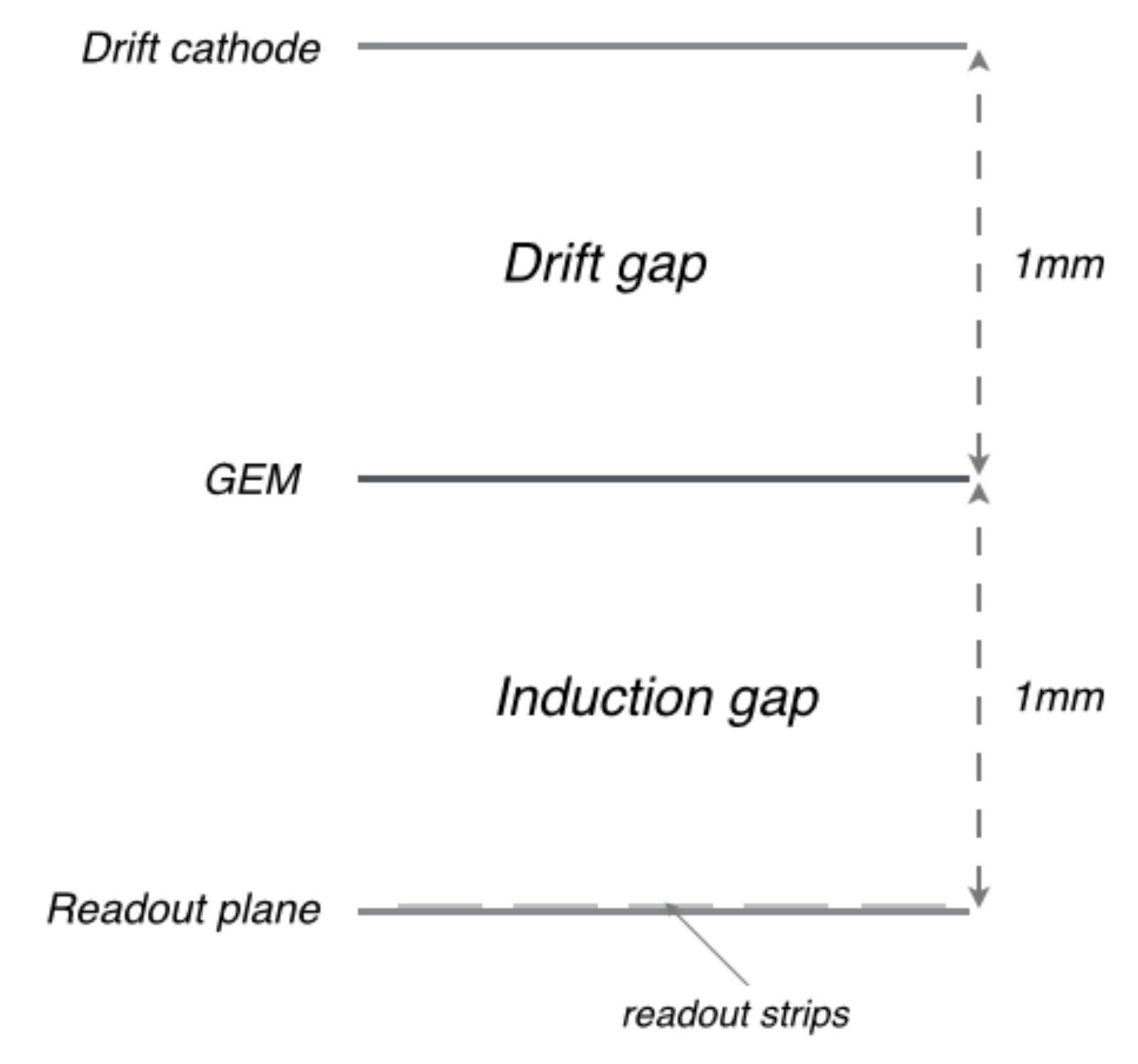}}
  \resizebox{7.25cm}{6.3cm}{\includegraphics{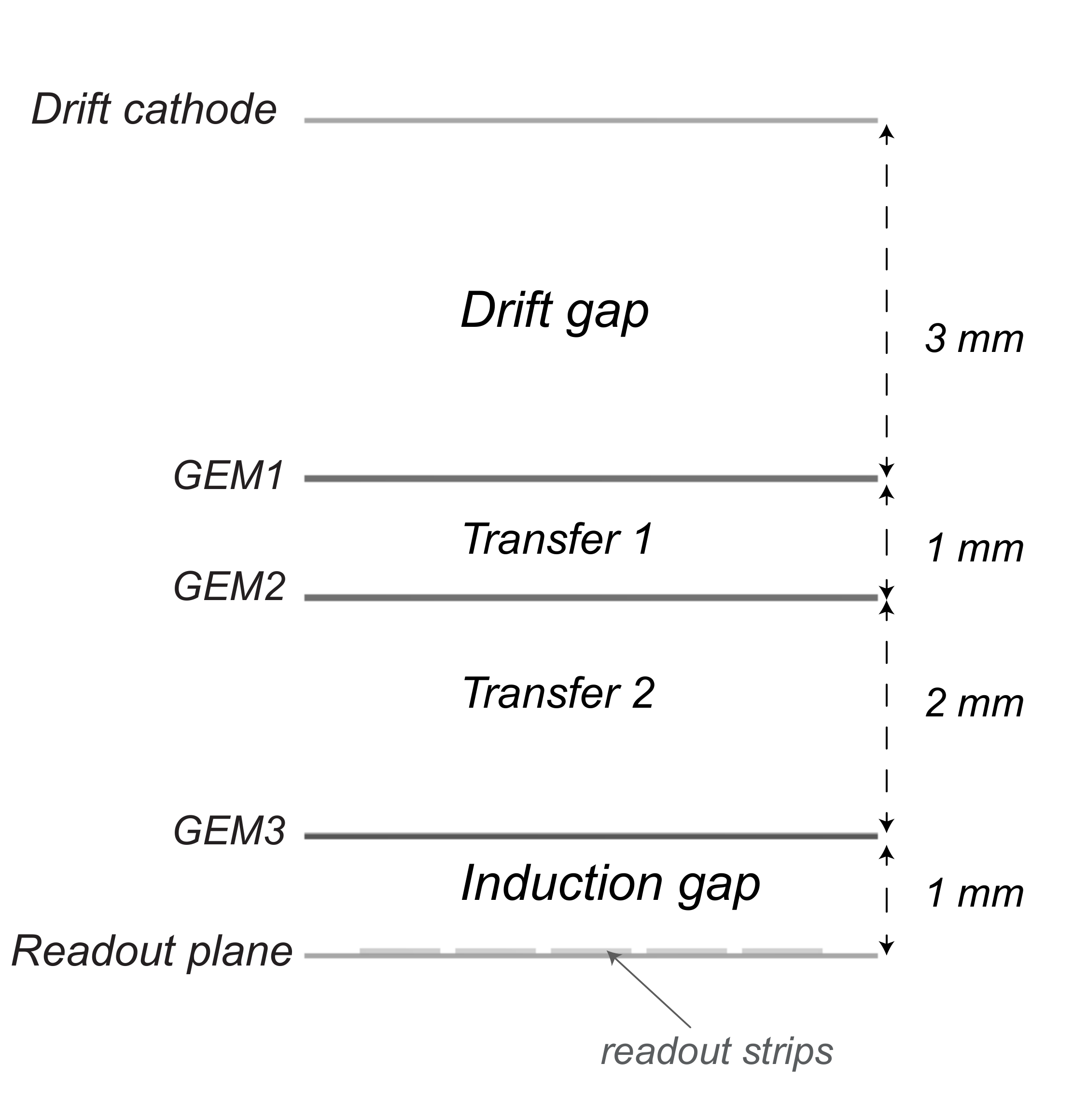}} \\
  \caption{Illustration of a single-GEM (left) and triple-GEM (right) detector.}
  \label{fig:gemgeometry}
\end{figure}

As the name suggests, a single-GEM detector consists of only one GEM foil. For this study, the distance between the drift cathode and the GEM foil as well as the GEM foil and the readout plane for a single-GEM detector is considered to be 1 mm each, as shown in figure~\ref{fig:gemgeometry} (left). The voltage is applied across the GEM foil as well as at the drift cathode while the readout plane is grounded.

A triple-GEM detector consists of a stack of three GEM foils placed at a relatively small distance from each other. The gas gaps in millimeters are 3:1:2:1, respectively, for the drift, transfer 1, transfer 2 and induction gaps as shown in figure~\ref{fig:gemgeometry} (right). The three GEM foils are labelled GEM1, GEM2 and GEM3 and voltage is applied across each of them. Also important is the voltage applied to the drift cathode, labelled as $V_{\rm Drift}$ later in this paper. These values have been optimized for CMS through a series of experimental and simulation studies~\cite{Gemtdr}. The readout strips are grounded and the gas mixture used inside the chamber is Argon (Ar) and Carbon dioxide (CO$_{2}$) with 70\% and 30\% proportions, respectively.

\subsection{Hole geometry}
\label{subsec:holegeometry}

\begin{figure}[hbtp]
  \begin{center}
    \resizebox{5.45cm}{6.3cm}{\includegraphics{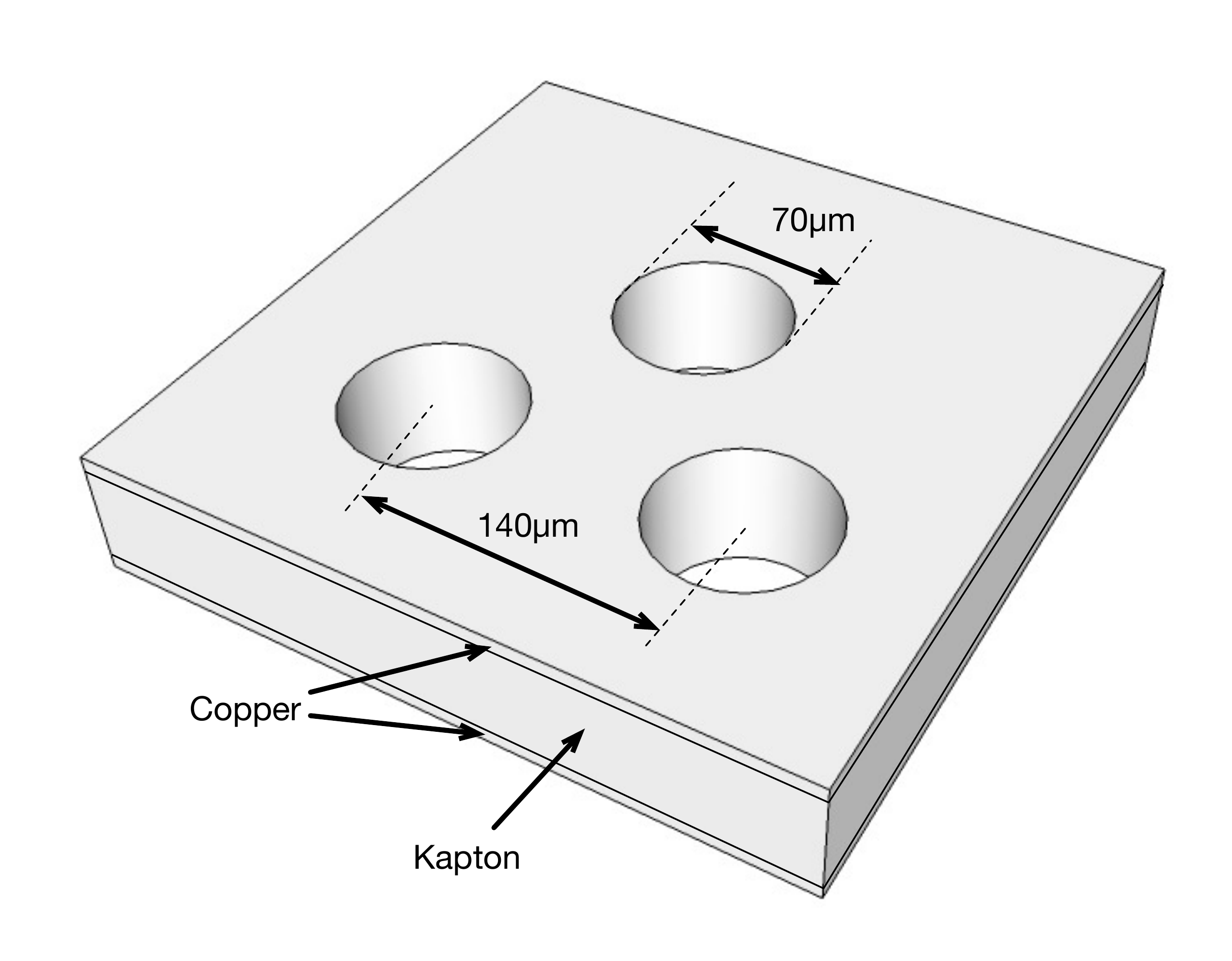}}
    \caption{Schematic view of a GEM foil with the standard hole pattern.}
    \label{fig:holes}
  \end{center}
\end{figure}

The GEM technology has been introduced as an amplification stage to overcome the problem of gas discharges in MPGDs~\cite{Sauli}. It consists of a 50 $\mu$m thick polymer foil clad on both sides with a 5~$\mu$m copper layer. The whole structure is chemically perforated with a high density of microscopic holes. These holes have double semi-conical shapes and the inner and outer diameter of the holes vary depending upon the etching procedure. A standard GEM foil has an inner diameter of 50~$\mu$m and an outer diameter of 70~$\mu$m for each hole which are spaced with a pitch of 140~$\mu$m in a hexagonal pattern (as shown in figure~\ref{fig:holes}).

\begin{figure}[hbtp]
  \begin{center}
 \resizebox{13cm}{7.2cm}{\includegraphics{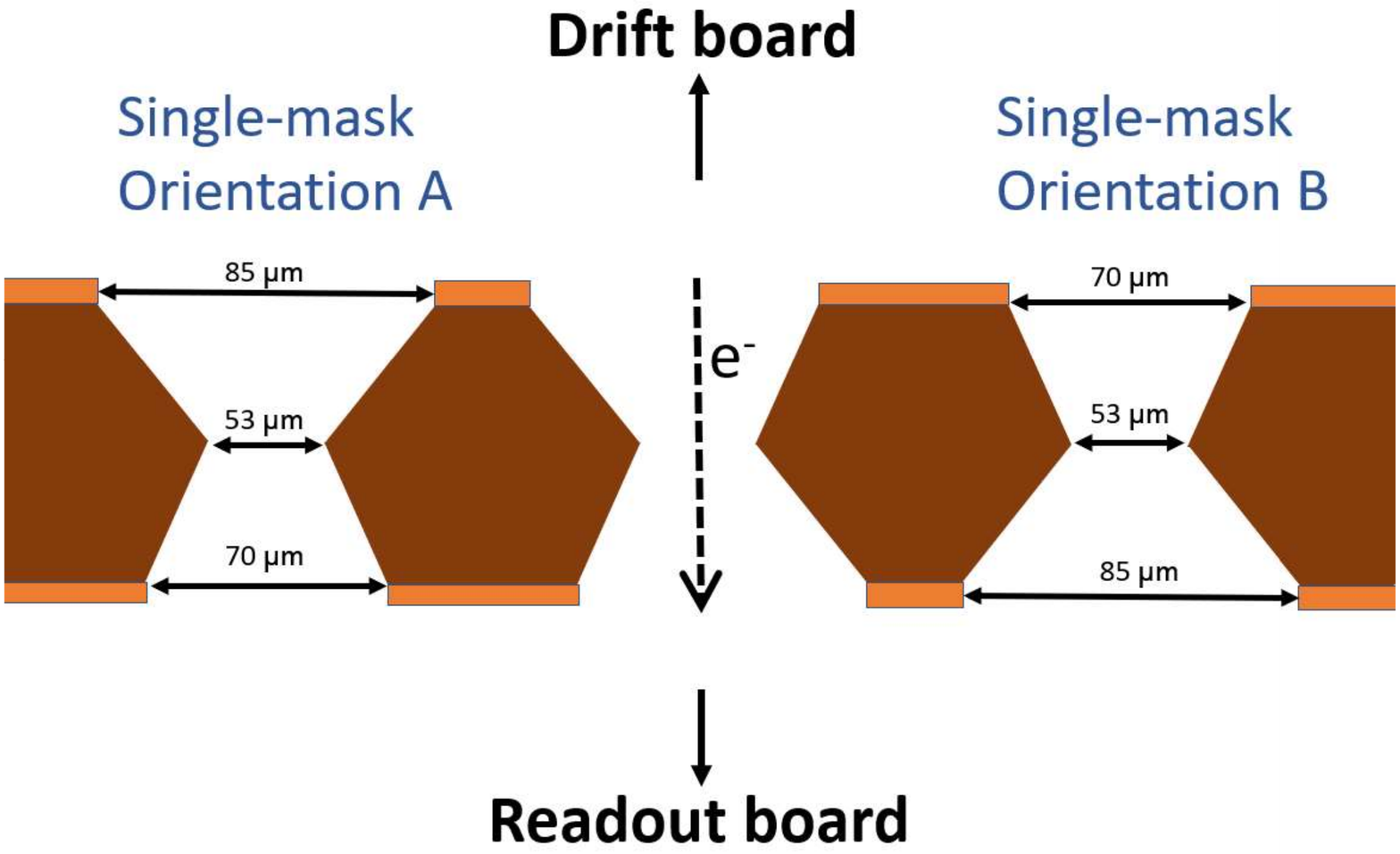}} 
 \caption{Cross sectional view for the two single-mask (SM) configurations. Left: orientation A with the larger hole diameter facing towards the drift cathode. Right: orientation B with the smaller hole diameter facing towards the drift cathode.}
 \label{fig:SM-OA-OB}
   \end{center}
\end{figure}

Depending on the etching technique two types of hole geometry result, i.e. double mask and single mask. In double mask etching, the Kapton layer of the GEM foil is equally etched from both sides whereas in single mask etching, the resultant hole diameters may be unequal on top and the bottom side of the foil. Since the holes in a single-mask GEM foil are not symmetrical on both sides, we define two orientations, depending on which side is facing the drift electrode. Orientation A (B) has the larger (smaller) diameter facing the drift electrode as shown in figure~\ref{fig:SM-OA-OB}.

Several hole geometries were considered depending on the foils used during the experimental studies. The foils were ordered from TECHTRA which were divided in three sets namely (i) Single mask, (ii) Double mask with asymmetric holes, and (iii) Double mask with symmetric holes.
The nominal diameters are around 85-50-70 for asymmetric foils, and 70-50-70 for symmetric foils. The production precision is $\pm$3$\mu$m on those nominal values.
One triple-GEM detector needs three foils, GEM1, GEM2 and GEM3 as defined in figure~\ref{fig:gemgeometry} (right). 
We combined only foils of the same type, either
\begin{description}
    \item (i) asymmetric foils from single-mask production with average hole diameters of 85-53-70, 
    \item (ii) asymmetric foils from double-mask production with average hole diameters of 85-50-70, and 
    \item (iii) symmetric foils from double-mask production with average hole diameters of 70-50-70.
\end{description}

The GEM chambers with asymmetric foils (configurations (i) and (ii)) were used with two orientations (A and B), as defined in figure~\ref{fig:SM-OA-OB} while keeping the same orientation of the foils inside the detector. Thus, there were a total of five different triple-GEM detector configurations obtained from these three sets of hole configurations. Similarly for single-GEM detector, five different configurations were obtained using the foil parameters quoted above.

\subsection{Voltage Settings}

A single-GEM detector can be operated by setting three voltages, two for the GEM foil (top and bottom) and the $V_{\rm Drift}$. Table~\ref{table:SingleGEMVolt} shows the voltages used during the simulation of the effective gain.

A triple-GEM detector has seven voltages to be set, two for each of the three GEM foils and the $V_{\rm Drift}$. The latter is usually given in the results graphs as the main quantity to determine the effective gain. By using the voltage divider which is shown in figure~\ref{fig:ExpSetup}, one actively sets only the highest negative voltage (= $V_{\rm Drift}$) while the remaining six voltages are tuned according to the distribution of resistors of the voltage divider. The readout plane is grounded. For the simulation, the voltages are derived from the experimental settings and summarized in table~\ref{table:TripleGEMvolt}.

\begin{table}[htb]
\caption{Voltages applied at drift and across the GEM foil for a single-GEM detector for the simulation study.}
\label{table:SingleGEMVolt}
\begin{center}
\begin{tabular}{|c|c|c|c|}
\hline
$V_{\rm Drift}$ (V) & $V_{\rm GEM-top}$ (V) & $V_{\rm GEM-bottom}$ (V) & $\bigtriangleup V_{\rm GEM}$ (V)\\ 
\hline
890 & 750 & 400 & 350\\ %\hline
940 & 800 & 400 & 400\\ %\hline
990 & 850 & 400 & 450\\ %\hline
1040 & 900 & 400 & 500\\ \hline
\end{tabular}
\end{center}
\end{table}

\begin{table}[htb]
\caption{Electric field across the drift, transfer 1, transfer 2, and induction gaps along with the potential difference across all three GEM foils in a triple-GEM detector for the simulation study.
%Voltages applied at drift and across the three GEM foils for a triple-GEM detector for the simulation study. Each GEM foil has two sides, here labeled as GEM1T for GEM1-top and GEM1B for GEM1-bottom.
}
\label{table:TripleGEMvolt}
\begin{center}
\begin{tabular}{|c|c|c|c|c|c|c|}
\hline
$E_{\rm Drift}$ & $\bigtriangleup V_{\rm GEM1}$  & $E_{\rm Transfer1}$ & $\bigtriangleup V_{\rm GEM2}$ & $E_{\rm Transfer2}$  & $\bigtriangleup V_{\rm GEM3}$  & $E_{\rm Induction}$\\ 
(V/cm) & (V) &  (V/cm) & (V) & (V/cm) & (V) & (V/cm)\\
%\multicolumn{5}{|c|}{Absolute Voltage (V)}\\ 
 \hline
% \hline
1993.33 & 300 & 2330 & 292 & 2325 & 280 & 3320\\ %\hline
2153.33 & 323 & 2520 & 316 & 2515 & 301 & 3590\\ %\hline
2313.33 & 347 & 2710 & 339 & 2700 & 323 & 3860\\ %\hline
2473.33 & 371 & 2890 & 363 & 2885 & 346 & 4120\\ %\hline
2713.33 & 407 & 3170 & 397 & 3165 & 380 & 4520\\ %\hline
%GEM3T & 1150 & 612  & 660  & 709  & 758  & 832 \\ %\hline
%GEM3B & 625  & 332  & 359  & 386  & 412  & 452\\ 
\hline
\end{tabular}
\end{center}
\end{table}

%%%%%%%%%%%%%%%%%%%%%%%%%%%%%%%%%%%%%%%%%%%%%%%%%%%%%%%%%%%%

\section{Measurements of different foil configurations in triple-GEM detectors}
\label{sec:ExpMeas}

\begin{figure}[htbp]
\centering
\resizebox{1.0\linewidth}{0.8\linewidth}{\includegraphics{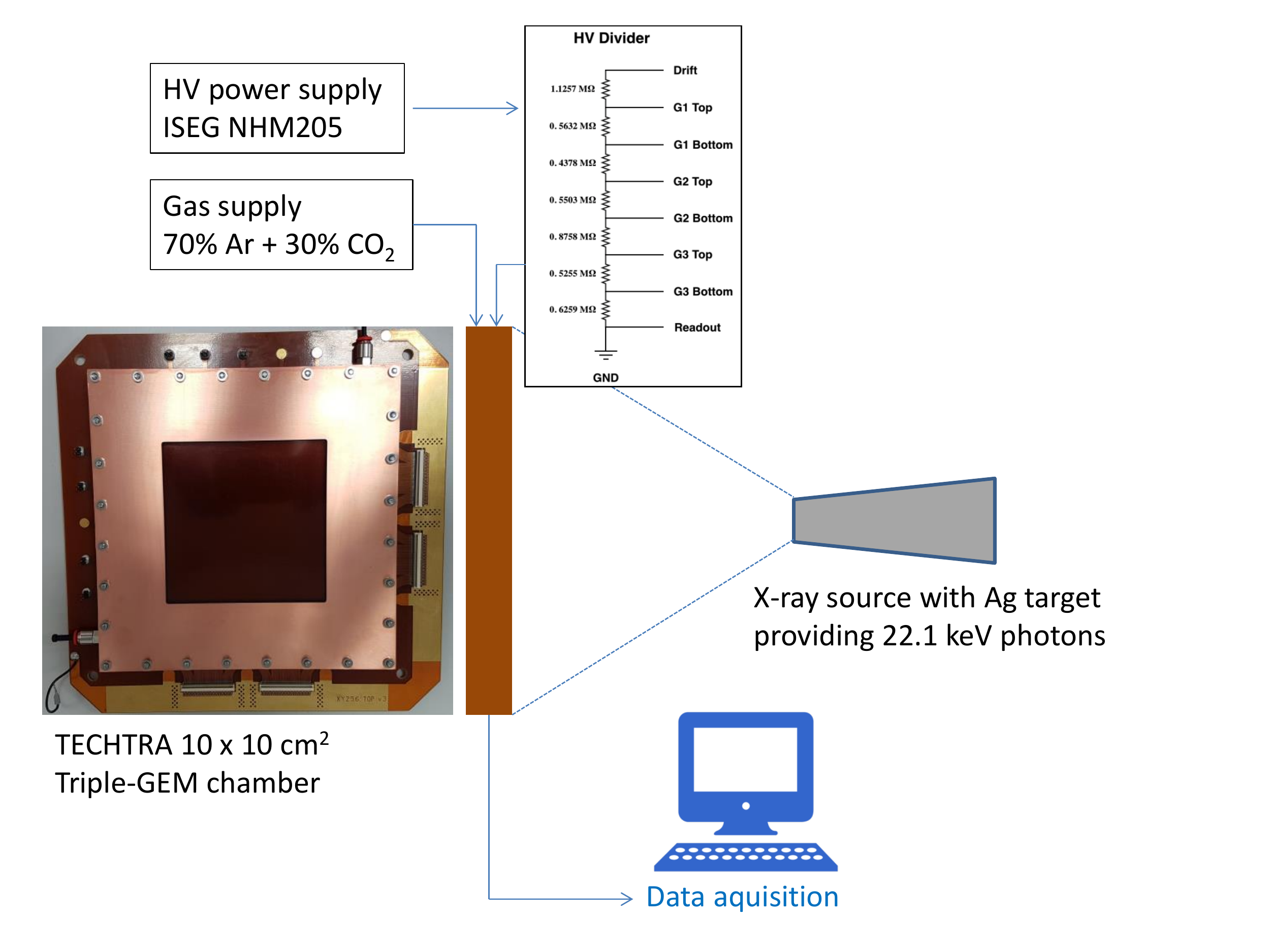}}
\caption{Schematic view of the experimental setup. The resistor values of the voltage divider are also given in table~\ref{table:TripleGEMvolt}.}
\label{fig:ExpSetup}
\end{figure}

This section provides details about the experimental set-up being used at Aachen, Germany to confirm the results obtained via simulations. The active area of GEM foils used for this study is 10 cm $\times$ 10 cm with both double-mask and single-mask etching having a pitch of 140 $\mu$m. All the triple-GEM detectors under study were built with 3/1/2/1 mm gap configuration (as defined in figure~\ref{fig:gemgeometry} (right)). The experimental setup is sketched in figure~\ref{fig:ExpSetup}. All three foils were powered up using a High Voltage (HV) resistive divider whose schematics is also shown. The main source of particles was a collimated beam of soft X-rays. The X-ray generator has a Ag-target yielding X-rays with a mean energy of 22.1~keV. 
These photons are mainly absorbed in the copper layer of the drift electrode and generate 8~keV fluorescent copper photons which subsequently interact with the gas.
The gas mixture used in all the measurements was a mixture of Ar:CO$_2$ in 70:30 proportion with a flow rate of 5 $\rm L/h$.

The high voltage supply used was ISEQ NHM205 and the output was taken from the readout of the 128 strips. The signal obtained by merging 128 strips of the triple GEM readout board was sent to an ORTEC 142 PC pre-amplifier, then amplified using an ORTEC 474 timing filter amplifier and finally sent to a LECROY 623 discriminator with a 100 mV threshold. The rate was then measured using a custom-made Scaler \& Counter over the required period of time. The current was measured with the help of a KEITHLEY pico-ammeter and recorded with a Labview program via a GPIB interface.

As all gas detectors, GEM detectors have properties that change with environmental conditions, in particular temperature and pressure. While the former can be artificially controlled, the latter cannot. In fact, the measurements were taken at a constant temperature of 295~K, while the pressure varied. The simulations were performed matching the laboratory temperature and pressure conditions during the data taking, in order to avoid the introduction of approximated correction factors for the several output variables.

%%%%%%%%%%%%%%%%%%%%%%%%%%%%%%%%%%%%%%%%%%%%%%%%%%%%%%%%%%%%

\section{Results}
\label{sec:results}

In this section, the summary of the results obtained from simulation as well as measurements are provided. A scan of effective gain was performed using simulations (for both single and triple-GEM detectors) and measurements (only for triple-GEM detector). 
In both cases the detector configurations from section~\ref{subsec:holegeometry} were used.
The voltages applied at the drift cathode as well as across each GEM foil during the simulations and experimental studies are provided in tables~\ref{table:SingleGEMVolt} and~\ref{table:TripleGEMvolt} for single-GEM detector and triple-GEM detector respectively. Section~\ref{sec:SingleGEMResults} discuss only the simulation results obtained for various configurations of single-GEM detector. A direct comparison of the results from simulations and data for triple-GEM detector are presented in section~\ref{sec:comparison}. The results obtained for orientation A and orientation B are further discussed in section~\ref{sec:Explain}.

\subsection{Simulation results for a single-GEM detector}
\label{sec:SingleGEMResults}

\begin{figure}[hbtp]
\centering
%\hspace*{0.2cm}
\resizebox{9cm}{9cm}{\includegraphics{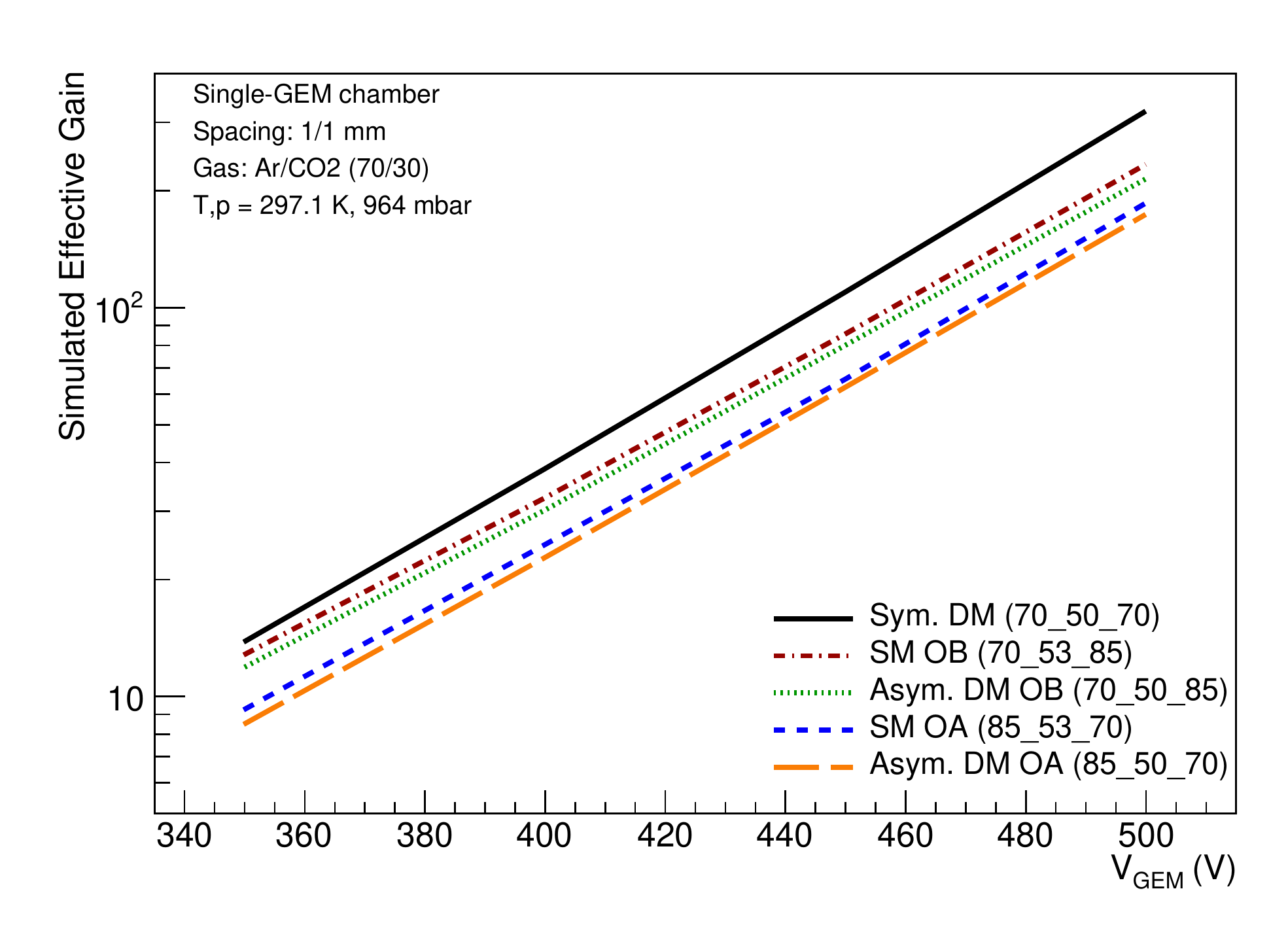}}
\caption{Simulated effective gain for a single-GEM detector as a function of voltage across the GEM foil for various configurations.}
\label{fig:SingleGEM}
\end{figure}

Using the method and tools described in section~\ref{sec:setup}, the effective gain of the GEM detectors has been simulated and studied in detail. The results for the gain variation in a single-GEM detector with respect to the voltage across the GEM foil are presented in figure~\ref{fig:SingleGEM} and an exponential behavior was observed for all the five configurations. It is also observed that orientation B always have higher gain as compared to orientation A, confirming the results in Ref.~\cite{JeremieThesis}, irrespective of the masking technique used.

\begin{figure}[hbtp]
\centering
%\hspace*{0.2cm}
\resizebox{7.2cm}{7.2cm}{\includegraphics{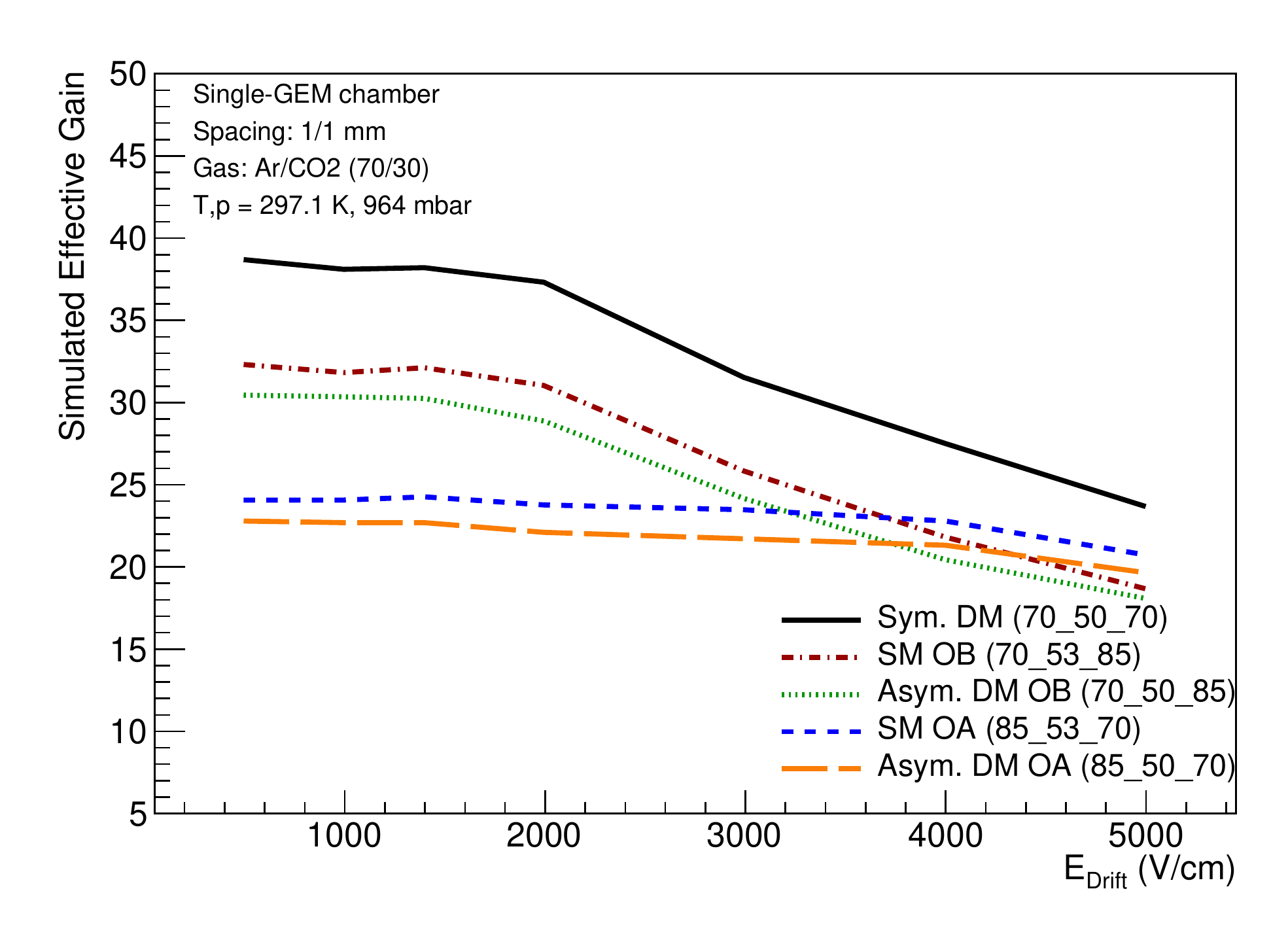}}
\resizebox{7.2cm}{7.2cm}{\includegraphics{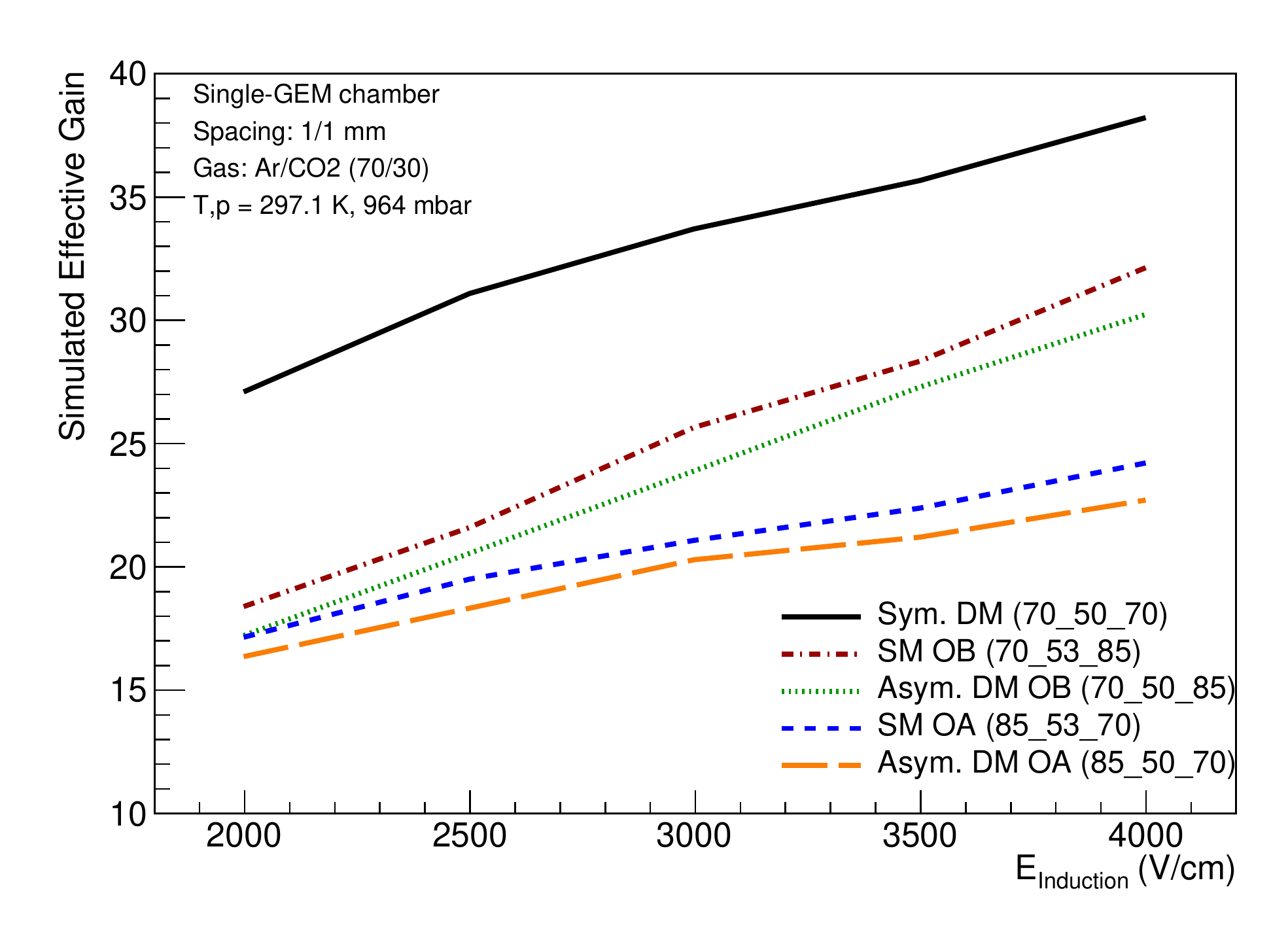}}
\caption{Left: simulated effective gain as a function of drift field for various configurations while keeping the voltage across the GEM foil as 400 V.
Right: simulated effective gain as a function of induction field for various configurations while keeping the drift field as 1400 V/cm.} 
\label{fig:DIVariation}
\end{figure}

The performance of a single-GEM detector was also tested while changing the drift field and the induction field of the detector. The effective gain was studied when varying the drift field from 500 V/cm to 5000 V/cm. This variation was obtained by changing the voltage at the drift plane ($V_{\rm Drift}$) but keeping the voltage across the GEM foil ($\bigtriangleup V_{\rm GEM}$) as constant i.e. 400 V. Figure~\ref{fig:DIVariation} (left) shows the variation of the effective gain with respect to the drift field and the results show that orientation A is less sensitive to the change in the electric field. Whereas, the detector with foil having orientation B and symmetric double mask shows a similar behavior i.e. almost constant gain until 2000 V/cm and then it decreases.

Another scan was performed by varying the induction field from 2000 V/cm to 4000 V/cm (in steps of 500 V/cm) while keeping the drift field as constant i.e. 1400 V/cm. Figure~\ref{fig:DIVariation} (right) shows that the effective gain always increases with an increase in the induction field although the rate of increase varies depending upon the hole configuration of the foil used.

\subsection{Experimental results with triple-GEM detectors compared to simulation}
\label{sec:comparison}

\begin{figure}[hbtp]
\centering
\resizebox{11cm}{10cm}{\includegraphics{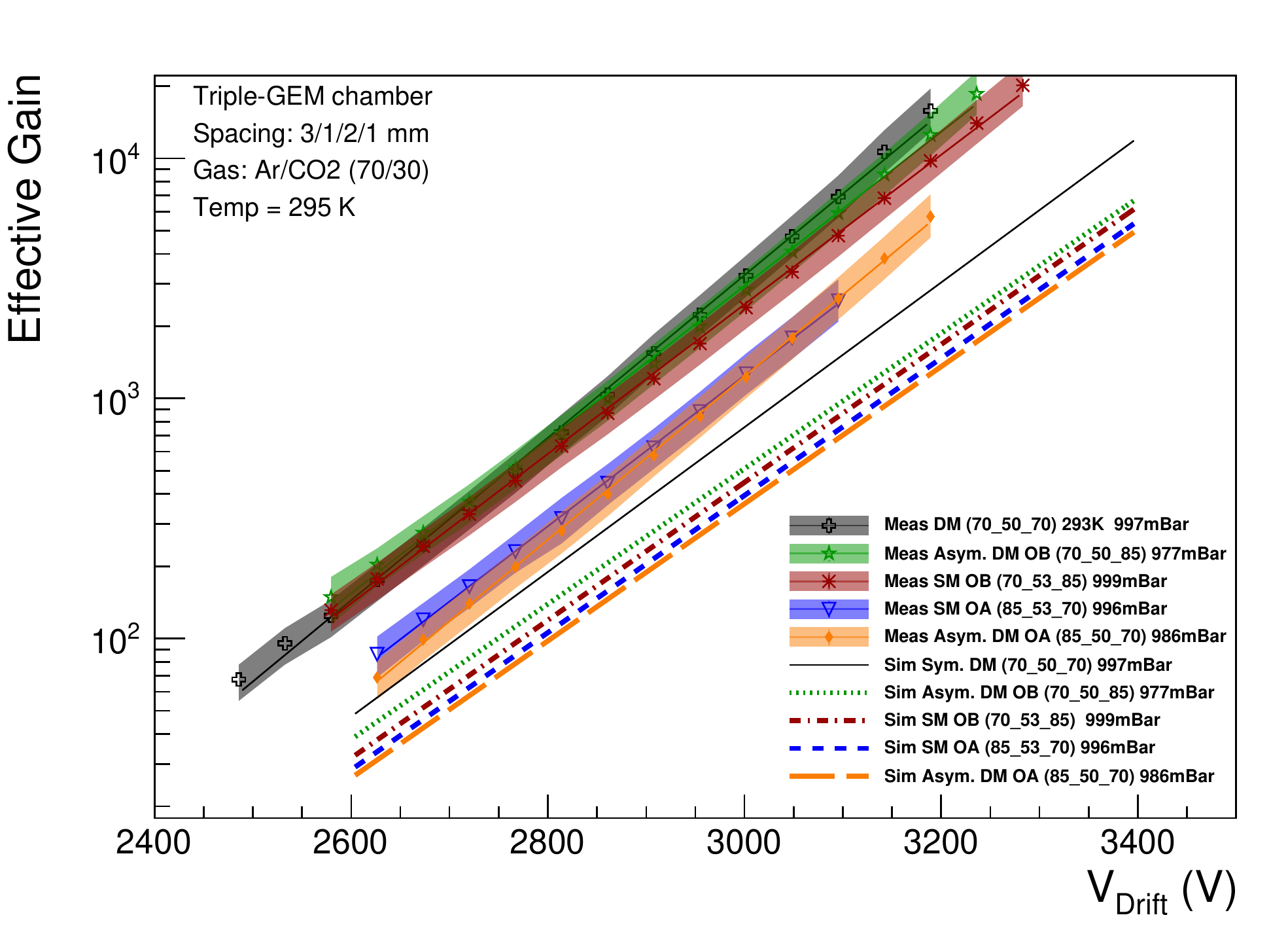}}
\caption{Measurements and simulation results of the effective gas gain of triple-GEM detector as a function of voltage across the detector for various configurations. The uncertainty band reflects variations from the $\pm$3$\rm \mu$m uncertainty of the real hole dimensions.}
\label{fig:ExpResult}
\end{figure}

\begin{figure}[hbtp]
\centering
\resizebox{7cm}{7cm}{\includegraphics{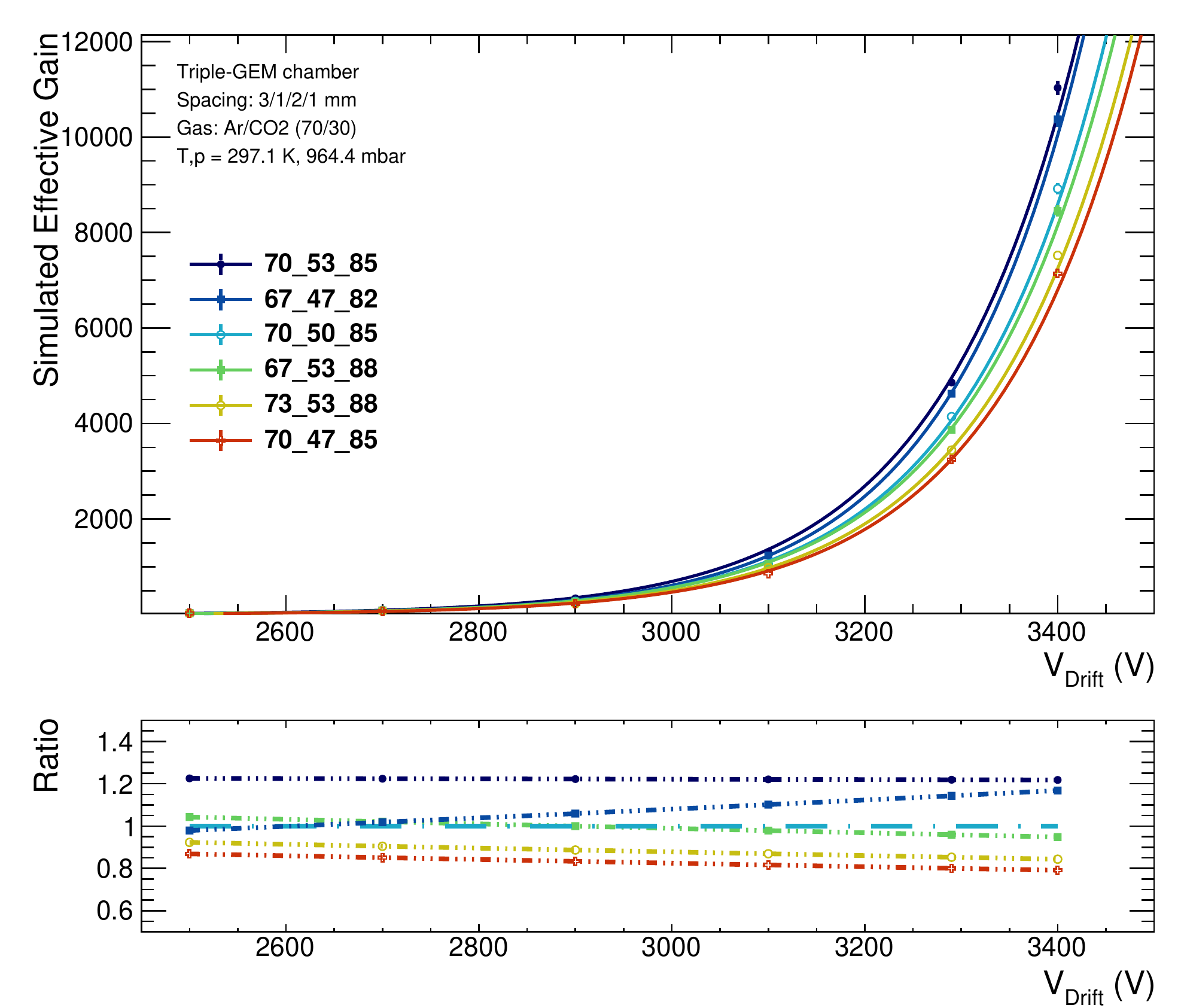}}
\resizebox{7cm}{7cm}{\includegraphics{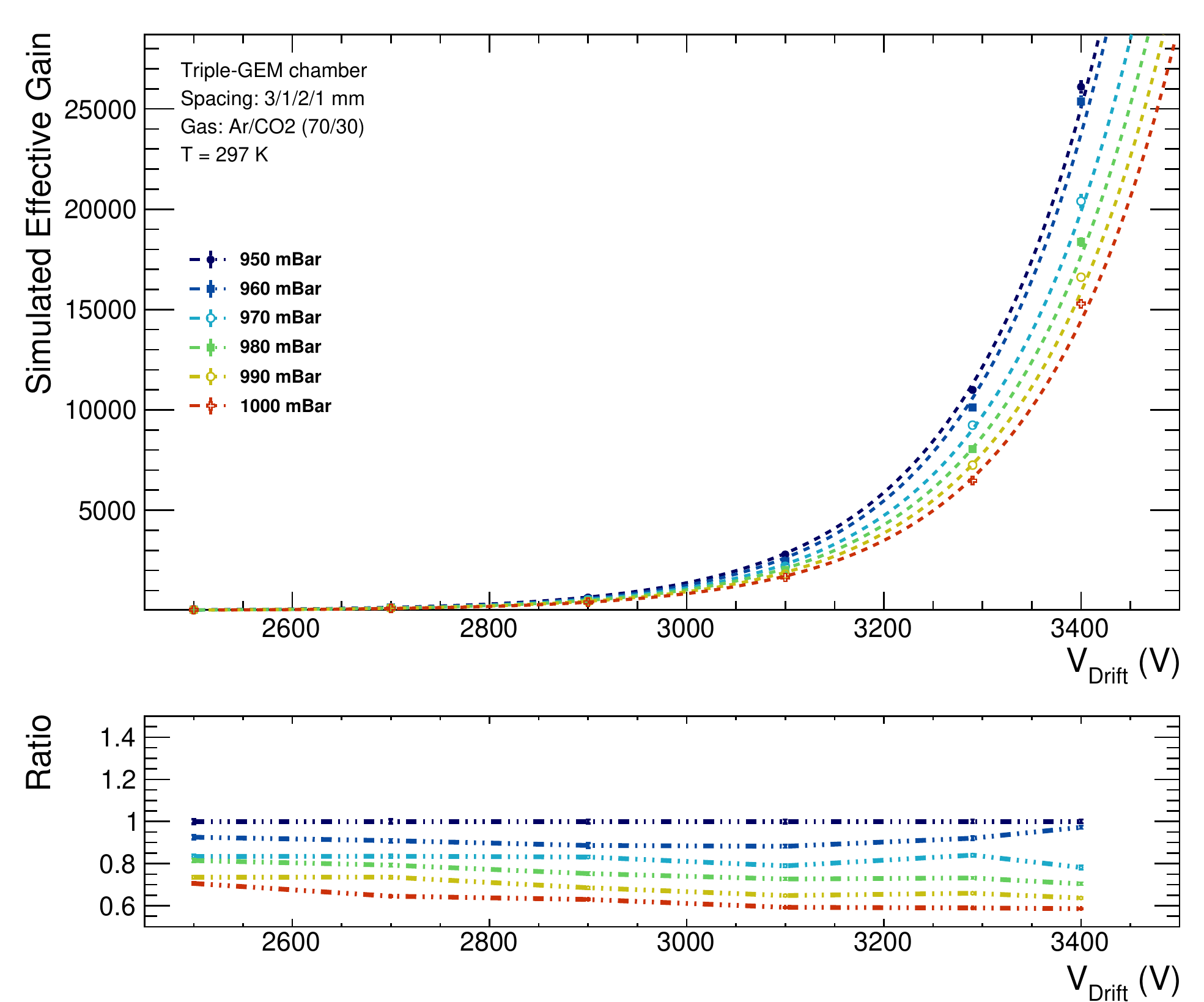}}
\caption{Left: simulation of varying hole diameters within the production accuracy. This simulation is the basis of the uncertainty bands in the previous figure. The lower panel shows the ratio of the studied diameters with respect to the reference of (70-50-85)~$\mu$m.
Right: impact of the pressure variation on the simulated effective gain. The lower panel shows the ratio with respect to the lowest pressure of 950~mbar corresponding to the highest gain.}
\label{fig:Variation}
\end{figure}

The effective gas gain of the triple-GEM detectors has been measured using the set-up explained in section~\ref{sec:ExpMeas}. The four asymmetric and the symmetric configuration 70-50-70 are measured. All these plus a second symmetric configuration (85-50-85)
were simulated using the nominal hole diameters.
In reality the hole diameters exhibit slight variations around the nominal values which are 85-50-70 or 70-50-85. The accuracy of production is given by the producer as $\pm$3$\mu$m. To understand the impact of the hole fluctuations on the measured gas gain, a range of parameters within the production accuracy was simulated
(see figure~\ref{fig:Variation} (left)). Based on that distribution an uncertainty band for each measurement of the four asymmetric configurations was extracted, shown in figure~\ref{fig:ExpResult}.

The effective gas gain for the measured GEM detector configurations as a function of the voltage applied to the drift cathode, $V_{\rm Drift}$, is depicted in figure~\ref{fig:ExpResult}. The comparison is performed using the real temperature and pressure conditions to ensure a fair comparison. All the detectors exhibit the expected exponential increase in the gain with increasing voltage. 
The measurements are shown with a small band around the measured points (shown as icons) representing the fluctuations due to the production accuracy. The corresponding simulations are performed for the nominal hole diameters and hence represented by only one point (shown by different line styles). Figure~\ref{fig:Variation} (right) shows the impact of varying the pressure between 950 and 1000 mbar. 

The triple-GEM detector built in orientation B, where the larger hole diameter faces the readout board, from either single mask or double mask technology with asymmetric holes are having approximately the same gas gain. Similarly, the triple-GEM detector built in orientation A from either single mask or double mask foils with asymmetric hole configurations are also having comparable gas gain. One can conclude that the orientation of the asymmetric hole is the key parameter, independently on the production mechanism with single or double mask. The effective gain for orientation B, where the larger diameter faces the readout board, is always substantially higher than orientation A. The symmetric double mask configuration with 70-50-70 shows the highest gas gain, which is expected given the narrow hole diameter both at top and bottom of the GEM foil.

\begin{figure}[hbtp]
\centering
%\hspace*{0.2cm}
\resizebox{9cm}{8cm}{\includegraphics{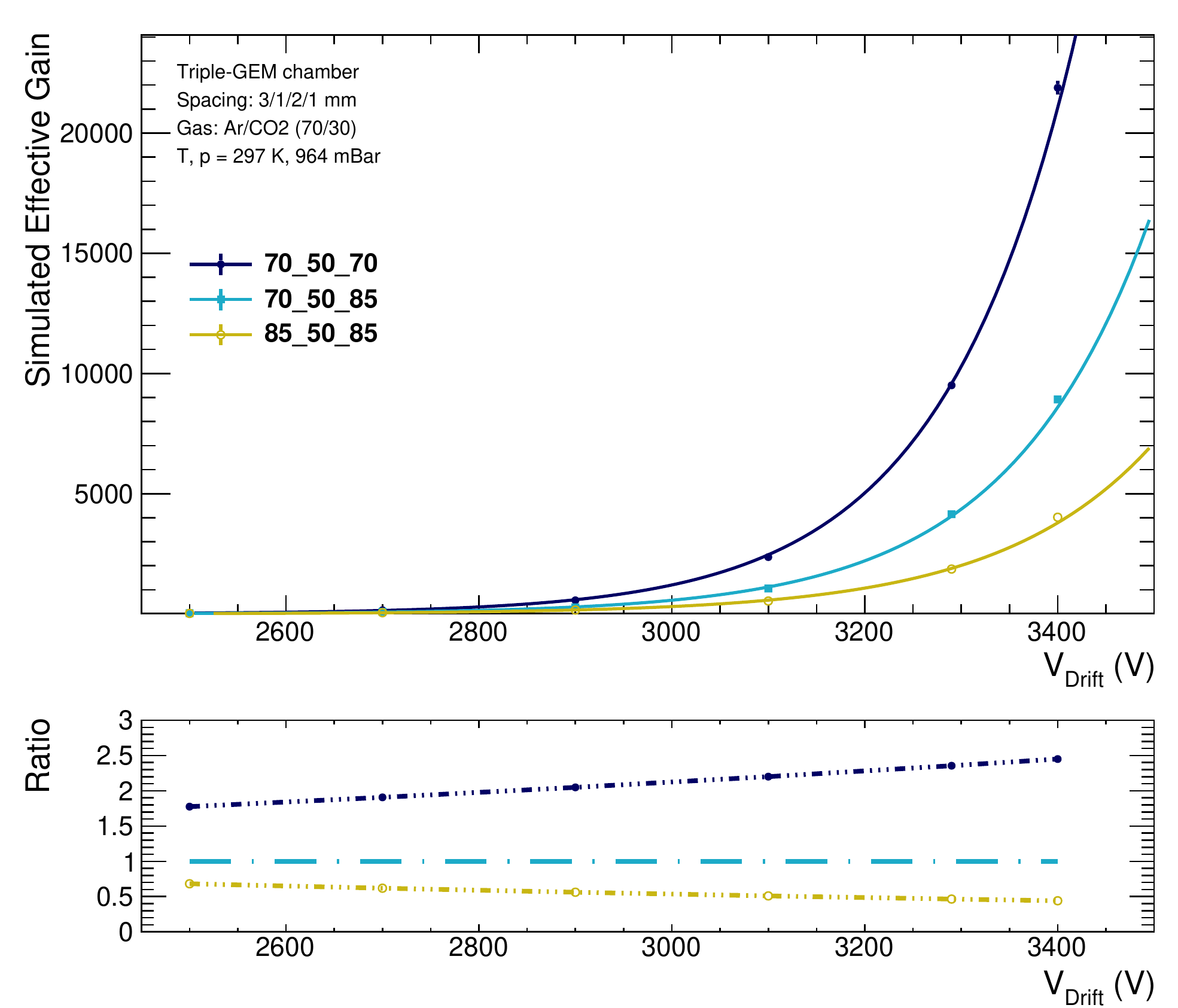}}
\caption{Simulated gas gain of triple-GEM detectors as a function of voltage across the detector for various configurations. The lower panel shows the ratio with respect to the reference of (70-50-85)~$\mu$m.}
\label{fig:Sim-symmetric}
\end{figure}

Figure~\ref{fig:Sim-symmetric} shows the simulated effective gain for the two symmetric configurations (70-50-70 and 85-50-85) along with one representative asymmetric configuration. The symmetric configuration 70-50-70 reflecting the smallest hole diameters clearly exhibit the largest gain. This is expected since the field strength 
goes as the inverse square of the hole diameter. The symmetric configuration 85-50-85 shows the expected lower gain in comparison to the 70-50-85 configuration. Only the configuration 70-50-70 was available for measurements and is included in the experimental results in figure~\ref{fig:ExpResult}. This detector configuration built with foils from double mask technology with symmetric holes, shows a gain comparable to orientation B
within the uncertainties. Given that the upper hole diameter is comparable to the one of orientation B (while the lower hole diameter is smaller), this demonstrates again the importance of the upper hole diameter.

\subsection{Explaining the observations}
\label{sec:Explain}

\begin{figure}[hbtp]
\centering
\resizebox{8cm}{7cm}{\includegraphics{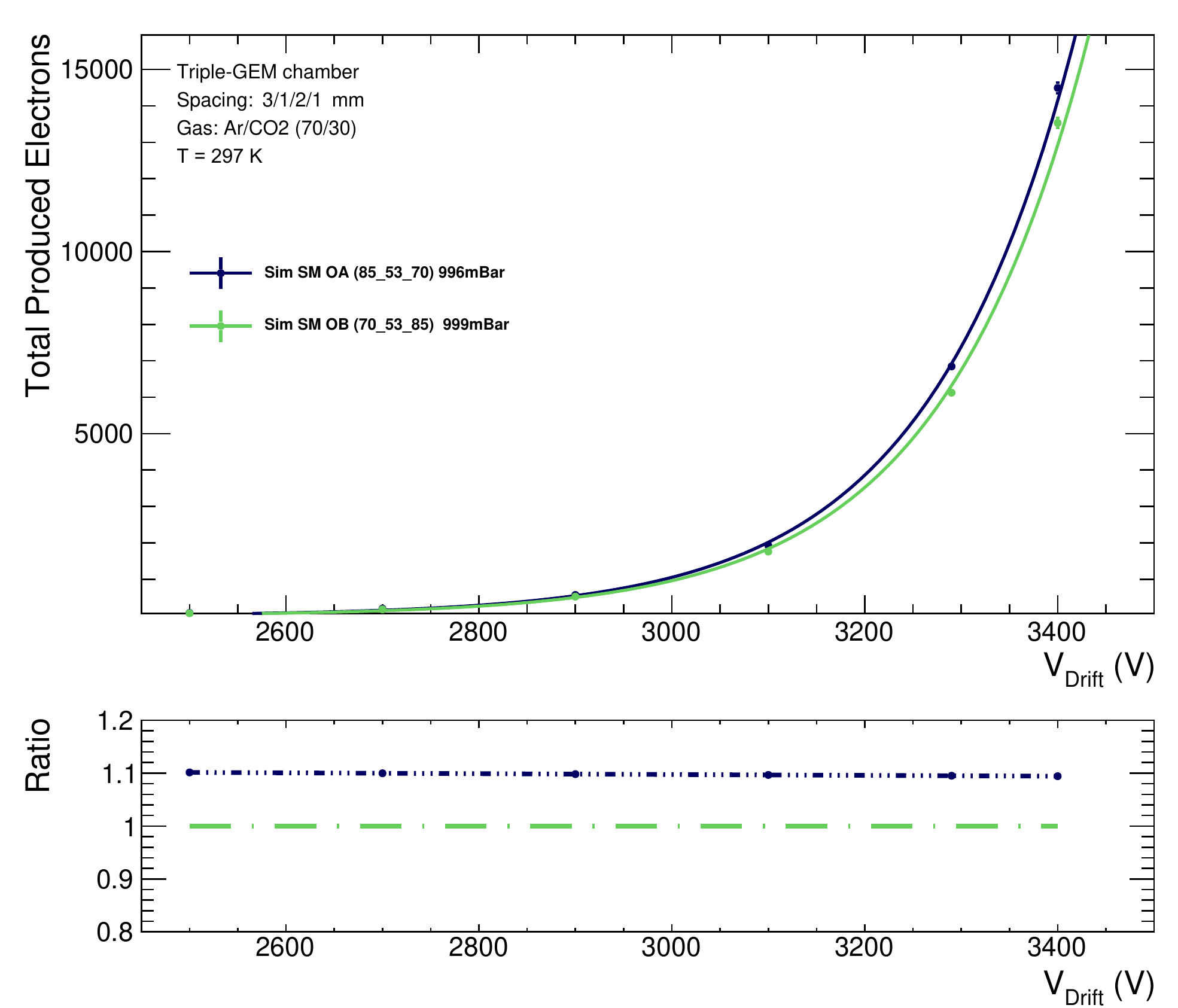}} 
\caption{Total number of electrons for the two single mask asymmetric foil geometries. The ratio in the lower panel shows a constant difference of about 10\% independent of the $V_{\rm Drift}$.}
\label{fig:TotEle}
\end{figure}

\begin{figure}[hbtp]
\centering
\resizebox{16cm}{8cm}{\includegraphics{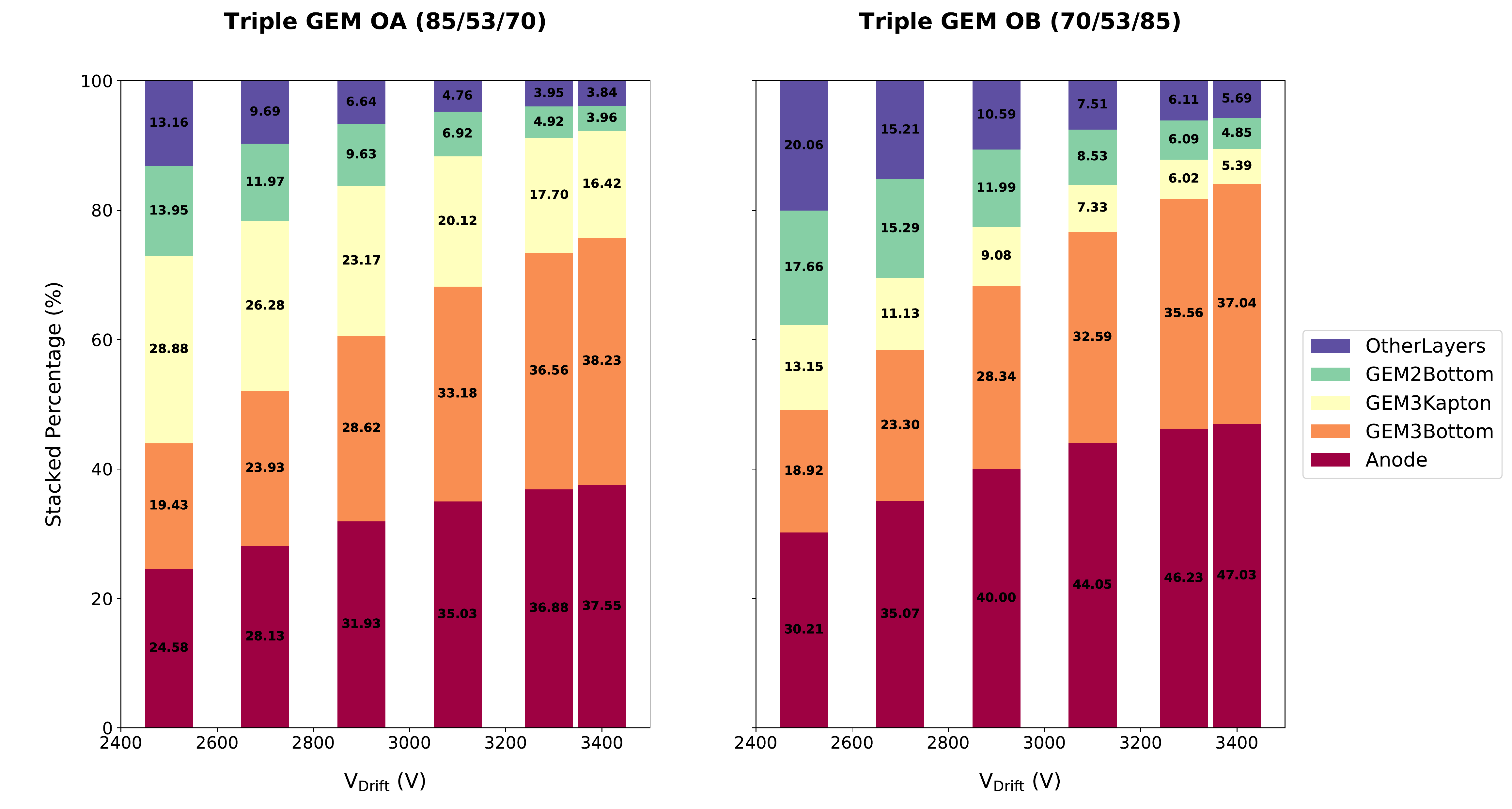}}
\caption{Stacked percentage of the stages reached by electrons; Left: for triple GEM with orientation A, Right: for triple GEM with orientation B. Reaching the anode yields a signal, while electrons stopping at the other stages are lost.}
\label{fig:EleLost}
\end{figure}

Studies are performed to understand why orientation A yields a lower gas gain. The simulations are carried out at a standard temperature and pressure to compare variables such as ``Fraction lost" and ``Total produced electrons". When comparing orientation A (85-53-70) with orientation B (70-53-85), the total number of produced electrons turns out to be comparable (see figure~\ref{fig:TotEle}). One has to study the various stages of electron loss to see a difference. Figure~\ref{fig:EleLost} shows the stacked percentage of the electrons reach, either ending at the anode (red), at GEM3 bottom (orange) and GEM3 Kapton (yellow), at GEM2 bottom (green) or any other GEM layers (blue). The latter are minor contributions and hence summed up. Only electrons reaching the anode contribute to the signal and this fraction is always larger in case of orientation B. Depending on the drift voltage between 30 and 48\% of electrons reach the anode for orientation B while only 24 to 37\% reach the anode for orientation A, hence the lower effective gain for the latter. Electrons ending at other stages are lost. The fraction of electrons reaching the anode increases with the drift voltage resulting in a higher effective gain.

The losses at GEM3 (bottom and Kapton) are the main loss-mechanism. It is also increasing with the drift voltage at the expense of contributions from other minor stages decreasing accordingly. For orientation B the sum of both GEM3 losses ranges from 30 to 42\% while the same loss ranges from 50 to 55\% for orientation A leading subsequently to the lower gain.

%%%%%%%%%%%%%%%%%%%%%%%%%%%%%%%%%%%%%%%%%%%%%%%%%%%%%%%%%%%%

\section{Summary}

Simulation studies and measurements were performed for the triple-GEM detector configuration. The foils used for these detector configurations have different hole geometries for single mask as well as double mask production technology. The variation in hole geometry clearly affects the gain of the GEM detectors as discussed above. For this study, the gap configuration was considered to be 3:1:2:1 (mm) for the drift, transfer 1, transfer 2 and induction gaps respectively. Comparison of the simulated gas gain of the triple-GEM detectors with orientation A and orientation B shows good agreement with the measurements. Orientation B exhibits a higher gas gain due to a smaller loss of electrons at GEM3 foil compared to orientation A, which is independent of the production technology as well as the detector geometry.

The total gas gain shows well-known discrepancies between measurement and simulation. The authors believe that this can be accounted due to some physical phenomena (like charging up, polarization, etc.) occurring inside the detector which were not considered during the simulation studies. Further, we would like to extend our study by including the charging up phenomenon since this effect can vary the gain of the detector to a much greater extent.

%\section*{Acknowledgements}
\acknowledgments{The authors wish to thank the outstanding support of R. Veenhof and H. Schindler. The work is supported by the Qatar National Research Fund under project NPRP 9-328-1-066.
The simulation work was performed using the High Performance Computing cluster of Texas A\&M University at Qatar.
This work was supported by the German Federal Ministry of Education and Research (grant no. 05H15PACC9 and grant no. 05H19PACC9). T. Kamon is supported in part by the U.S. Department of Energy grant de-sc0010813.}

%------------------------------------------------------------------------------

\end{document}